\documentclass[11pt]{article}

\usepackage{aas_macros,amsmath,amssymb,comment,cite,esint,graphicx,mathtools}
\usepackage[margin=.8in,letterpaper]{geometry}
\usepackage[colorlinks=true]{hyperref}
\usepackage[affil-it]{authblk}
\usepackage{subcaption}
\usepackage[utf8]{inputenc}
\usepackage{mathrsfs}
\usepackage{appendix}
\usepackage{amssymb}
\usepackage{float}
\usepackage{color}
\usepackage{cite}
\usepackage{hyperref}
\hypersetup{pageanchor=false}
\usepackage{indentfirst}
\usepackage{url}
\usepackage{float}
\usepackage{caption}
\usepackage[numbers,square,comma,sort&compress,merge]{natbib}
\usepackage{esint}
\usepackage{overpic}
\usepackage{graphicx}
\usepackage{epsf,amsmath,bbold,amsfonts,stmaryrd}
\usepackage{textcomp}
\usepackage{ulem}
\usepackage{tikz}

\numberwithin{equation}{section}
\let\originalleft\left
\let\originalright\right
\renewcommand{\left}{\mathopen{}\mathclose\bgroup\originalleft}
\renewcommand{\right}{\aftergroup\egroup\originalright}
\mathcode`\*="8000
{\catcode`\*=\active\gdef*{\mathclose{}\,\mathopen{}}}

\newcommand{\be}{\begin{equation}}
\newcommand{\ee}{\end{equation}}
\newcommand{\bea}{\setlength\arraycolsep{2pt} \begin{eqnarray}}
\newcommand{\eea}{\end{eqnarray}}
\newcommand{\nn}{\nonumber}

\newcommand{\cir}{{\text{IF}}}
\newcommand{\EM}{{\text{EM}}}
\newcommand{\ion}{{\text{ion}}}
\newcommand{\te}{{\text{e}}}
\newcommand{\tc}{{\text{c}}}
\newcommand{\mT}{{\mathcal{T}}}

\def\a{\alpha}

\def\D{\Delta}
\def\f{\frac}

\def\m{\mu} 
\def\n{\nu} 
\def\nn{\nonumber}
\def\pl{\partial}
\def\p{\phi}

\def\s{\sigma}
\def\S{\Sigma}
\def\t{\theta}
\def\T{\Theta}

\def\o{\omega}

\def\be{\begin{equation}}
\def\ee{\end{equation}}
\def\bag{\begin{aligned}}
\def\eag{\end{aligned}}
\def\bea{\begin{eqnarray}}
\def\eea{\end{eqnarray}}
\def\ba{\begin{array}}
\def\ea{\end{array}}

\def\bc{\begin{center}}
\def\ec{\end{center}}

\begin{document}
\title{A new analytical model of magnetofluids surrounding rotating black holes}
	
\author{Yehui Hou$^{1}$, Zhenyu Zhang$^{1}$, Minyong Guo$^{2}$, Bin Chen$^{1, 3, 4\ast}$}
\date{}
	
\maketitle
\vspace{-10mm}

\begin{center}
{\it
$^1$Department of Physics, Peking University, No.5 Yiheyuan Rd, Beijing
100871, P.R. China\\\vspace{4mm}

$^2$ Department of Physics, Beijing Normal University,
Beijing 100875, P. R. China\\\vspace{4mm}

$^3$Center for High Energy Physics, Peking University,
No.5 Yiheyuan Rd, Beijing 100871, P. R. China\\\vspace{4mm}

$^4$ Collaborative Innovation Center of Quantum Matter,
No.5 Yiheyuan Rd, Beijing 100871, P. R. China\\\vspace{2mm}
}
\end{center}

\vspace{8mm}

\begin{abstract}
In this study, we develop a simplified magnetofluid model in the framework of GRMHD. We consider an ideal, adiabatic fluid composed of two components,  ions and electrons, having a constant ratio between their temperatures. The flows are assumed to be governed by gravity, enabling us to employ the ballistic approximation, treating the streamlines as timelike geodesics. We show that the model is analytically solvable around a rotating black hole if the angular velocity of the geodesic $u^\t$ is vanishing. In the corresponding solution, which is named the conical solution,  we derive a comprehensive set of explicit expressions for the thermodynamics and the associated magnetic field. Furthermore, we explore the potential applications of our model to  describe the thick disks and the jets at the horizon scale. Our model provides a direct pathway for the study of black hole imaging.
\end{abstract}

\vfill{\footnotesize $\ast$ Corresponding author: bchen01@pku.edu.cn}

\maketitle

\newpage
\baselineskip 18pt
\section{Introduction}\label{sec1}

The unveiling of horizon-scale images of black holes captured by the Event Horizon Telescope (EHT) has sparked widespread interest and fascination \cite{EventHorizonTelescope:2019dse, EventHorizonTelescope:2021srq, EventHorizonTelescope:2022wkp, Lu:2023bbn}. The integration of theoretical and observational findings enables us to attain a clearer comprehension of the intricate physics surrounding black holes \cite{Goddi:2016qax, Vagnozzi:2019apd, EventHorizonTelescope:2021srq,  EventHorizonTelescope:2021dqv, Vagnozzi:2022moj}. Numerous studies, including the efforts of the EHT collaboration, suggest that the millimeter-wave emissions  observed  originate from the accretion disk or the surface of the jet, namely the funnel wall jet (FWJ) \cite{Hawley:2005xs, McKinney:2006tf, Moscibrodzka:2015pda, Chael:2018gzl, Nakamura:2018htq, EventHorizonTelescope:2019pgp, Kawashima:2020rmr, Lu:2023bbn}. The observational signatures of the disks and FWJs depend not only on gravitational effects but also significantly on the intrinsic physical properties of the magnetofluid, including electron distribution, temperature, and the associated magnetic field \cite{EventHorizonTelescope:2019pgp}. Theoretically, it is strongly demanding for an analytical model in the framework of general relativistic magnetohydrodynamics (GRMHD),  which can effectively capture essential characteristics of the magnetofluid. Such a model would greatly facilitate the study of the magnetofluid surrounding the black holes and enhance the research on  black hole imaging.

There have been many analytical models on accretion disk. The Novikov-Thorne disk, an extension of the standard disk model \cite{shakura1973black}, was established as the typical geometrically thin, equatorial accretion disk model in a relativistic background \cite{page1974disk, thorne1974disk}. Other studies have also explored different thin disk models \cite{1976Relativistic, Narayan:1994xi, Compere:2017zkn, Mitra:2022iiv}. Generally, geometrically thin disks are optically thick and emit black body radiation. However, recent observations by the EHT and other studies \cite{EventHorizonTelescope:2019pgp, EventHorizonTelescope:2022urf} suggest that accretions close to super massive black holes may exhibit geometrically thick, optically thin structures due to the dominant influence of gravity, which hinders rapid cooling and compression of matter in the vertical direction \cite{Narayan:1994xi}. Theoretical investigations related to geometrically thick disks are still ongoing \cite{Abramowicz:2011xu}.  Abramowicz et al. proposed a torus model consisting of pure toroidal flows \cite{abramowicz1978relativistic}, which was subsequently extended to include magnetized tori \cite{Komissarov:2006nz}. The torus model has proven to be a suitable initial condition for GRMHD simulations. Additionally, by employing the ballistic approximation that considers only the geodesic motions of streamlines, off-equatorial streamlines can be analytically solved \cite{Ulrich:1976zz, huerta2007simple}. However, these studies primarily focus on the contribution of ballistic flows to the formation of thin disks \cite{Tejeda:2012kb, Tejeda:2019lie}, without considering horizon-scale magnetofluids. Besides, there are many other studies of the phenomenological models of accretion flows, with the aim of applying these models to the black hole images of EHT \cite{Broderick:2005jj, Broderick:2006ez, Broderick:2008qf, Broderick:2009ty, Pu:2016qak, Pu:2018ute, Vincent:2014erw, Vincent:2019nni}.

Most of the existing disk models have predominantly focused on circular orbits and aimed to capture the dynamics on large scales. However, the potential to obtain highly resolved images of black holes presents an opportunity to reveal the behaviors of fluid at the horizon scale. Consequently, investigating the fluid dynamics near black holes has become increasingly important. In \cite{Compere:2017zkn}, the authors examined the Novikov-Thorne disk around a high-spin black hole and discovered a self-similar solution in the near-horizon region. In a recent study \cite{Mummery:2023tgh}, the authors analyzed the equatorial accretion flow inspiraling from the innermost stable circular orbit (ISCO) of a Kerr black hole and derived analytical thermodynamic solutions. However, it should be noted that the accretion flow close to the black hole is more likely to be geometrically thick. Furthermore, the polarized images of M87$^\ast$ suggest a preference for poloidal alignment in the magnetic field \cite{EventHorizonTelescope:2021srq}, implying a 
non-ignorable poloidal velocity within the magnetofluid. Hence, it is necessary to investigate the dynamics and morphology of a geometrically thick disk with nonvanishing poloidal inflows.

For the study of relativistic jets, despite various theoretical models \cite{Blandford:1977ds, Blandford:1982di, Rueda:2022fgz}, the real launching mechanism remains unknown. The outer boundary of the jet (FWJ) in the launching region is composed of hot plasma, thus contributing to the millimeter-wave black hole images through synchrotron radiation. However, the dynamics of the FWJ are still uncertain, despite some theoretical proposals \cite{Blandford:1982di, Ptitsyna:2015nta, Bu:2018rcp, Li:2019sgx, Yang:2021syv}. 
	
With these questions in mind, we propose a self-consistent, analytical model for horizon-scale magnetofluids in the framework of GRMHD, trying to capture key features of thick disks and FWJs. Specifically, we focus on high-temperature systems with ultra-relativistic electrons in the adiabatic limit. To simplify the analysis, we introduce a linear relationship between the temperatures of electrons and ions, represented by a constant factor, $z =T_{\text{ion}}/T_{\text{e}}$, and consider both the non-relativistic and ultra-relativistic limits of ions. With the relativistic Euler equations for ideal, adiabatic magnetofluids, we express the temperature and pressure as power functions of the particle number density, with the exponents determined by $z$. Then, we demonstrate that in the scenarios where both the sound speed and Alf\'{v}en velocity are sub-relativistic, the streamlines follow geodesics in the leading-order  approximation, namely, the ballistic approximation. Furthermore, for a stationary, axisymmetric, ballistic fluid configuration in Kerr spacetime, we derive explicit expressions for the fluid thermodynamics under specific conditions regarding the streamlines. Especially, one of the conditions leads to the conical solution, which is of particular importance in the study of thick disks and FWJs.  We also determine the accompanying magnetic field structure in the conical solution under the ideal MHD condition. These self-consistent results enable the construction of thick disk and FWJ models that capture essential characteristics of the emission profiles of astrophysical black holes. It is worth mentioning that we also employ the astrophysical parameters of M87$^\ast$ to estimate the feasibility of the approximations used in our model.

The structure of this manuscript is outlined as follows: Sec. \ref{sec2} presents  the framework and introduces the fundamental equations of the ideal and adiabatic fluid.  In Sec. \ref{ballistic}, we introduce the ballistic approximation and examine its applicability to the case of M87$^\ast$. In Sec. \ref{theKerr}, we derive the explicit expressions for the thermodynamics in Kerr spacetimes. We investigate the conical solution of the streamlines in detail in Sec. \ref{Con}. The associated magnetic field is discussed in Sec. \ref{magnetic}. Moving on to Sec. \ref{sec4}, we apply the conical solution to the study of thick disks and FWJs, and use graphical illustrations to help understanding. We conclude in Sec. \ref{sec5} with summary and discussions. We include some technical details in a few appendices. Throughout this study, we adopt the unit with $G = c = 1$.
	
\section{Basic equations for ideal and adiabatic fluid}\label{sec2}

In magnetohydrodynamics (MHD), the stress-energy tensor comprises several components: the fluid part, the viscosity part, the Maxwell part, and the radiation part. Different magnetofluid models adopt different forms of the stress-energy tensor. For instance, thick disk models often focus solely on the fluid part, disregarding other components, while thin disk models assume a negligible Maxwell part. In this study, we assume that the viscosity and radiation parts of the stress-energy tensor are negligible, and we only consider the fluid and Maxwell parts. In other words, we neglect the effects of radiative losses and diffusion during the evolution of fluid dynamics.

In fact, we will study the MHD of an ideal fluid in the presence of a magnetic field. 
For an ideal fluid, its stress-energy tensor can be written as 
\bea\label{idf}
T^{\m\n}_\cir=u^{\m}u^{\n}\left(\Xi+p \right) + g^{\m\n}p\, , 
\eea
where $u^\m$ is the 4-velocity, $p$ is the isotropic pressure, $\Xi$ is the internal energy density of the fluid. For the electromagnetic field $F_{\mu\nu}$, it obeys the equation $F_{\mu\nu}u^\nu = 0$, which suggests the absence of an electric field as measured by the fluid. The Maxwell stress-energy tensor can be expressed as\cite{McKinney:2006tf}:
\bea
T^{\m\n}_{\EM} = -\f{1}{4}g^{\mu\nu}F^2+F^{\m\a}F^{\n}_{\a} 
= \bigg( u^\m u^\n + \f{1}{2}g^{\m\n} \bigg)B^2-B^\m B^\n \, ,
\eea
where $B_{\m}$ is the magnetic field measured by the fluid. Thus, the total stress-energy tensor can be written as
\bea\label{tst}
T^{\m\n}=T^{\m\n}_{\cir} +T^{\m\n}_{\EM}\,.
\eea

The fluid dynamics are governed by the conservation law, known as the relativistic Euler equations,
\bea\label{conlaw1}
\nabla_{\m}T^{\m\n}&=& 0 \,.
\eea
Integrating Eqs. (\ref{idf}), (\ref{tst}) and (\ref{conlaw1}), we can find
\bea\label{gt}
0=\nabla_{\m} T^{\m\n}_{EM}+\left(g^{\m\n}+u^\n u^\m\right)\nabla_\m p+\left(\Xi+p\right)u^\m\nabla_\m u^\n+P u^\n\,,
\eea
where we have introduced a scalar function $P=u^\m \nabla_\m \Xi+\left(\Xi+p\right)\nabla_\m u^\m$. Upon contracting Eq. (\ref{gt}) with $u_\n$, we read
\bea\label{P0}
P=u^\m \nabla_\m \Xi+\left(\Xi+p\right)\nabla_\m u^\m=0\,,
\eea
where we have used the relation $u_\m \nabla_\n T^{\m\n}_{\text{EM}} = 0$ \footnote{For a detailed exposition of the proof for this formula, one might refer to Appendix C in \cite{Mummery:2023tgh}.}.  Then, Eq. (\ref{gt}) can be simplified and reduced to 
\bea\label{dum}
u^\m \nabla_{\m}u^\n = -\f{1}{\Xi+p} \big[ (u^\m u^\n + g^{\m\n}) \nabla_\m p + \nabla_{\m} T^{\m\n}_{\EM}  \big]\,,
\eea
which governs the dynamics of $u^\m$ along the streamlines.

Furthermore, we consider the conservation law for the particle number
\bea\label{cl2}
\nabla_\m\left(nu^\m\right)=0\,,
\eea
where $n$ is the number density of the plasma.  As the most abundant gases in the universe are hydrogen and helium, in a fully ionized plasma, it is typically composed of negatively charged electrons and positively charged hydrogen or helium ions. For the sake of simplicity, let us assume that all the positive ions are hydrogen ions, thus the number densities  of electrons and ions are equal, $n_\te = n_\ion = n$. From Eq. (\ref{cl2}), we have
\bea\label{neq}
\nabla_\m u^\m=-\frac{u^\m\nabla_\m n}{n}=-\f{d}{d\tau}\log{n}\,,
\eea
where we have defined the proper time $\tau$ along the streamline such that $\f{d}{d\tau}=u^\m\nabla_\m$. And Eq. (\ref{P0}) can be rewritten as 
\bea\label{teq}
\f{d \Xi}{d\tau}=-\left(\Xi+p\right)\nabla_\m u^\m=\frac{\Xi+p}{n}\f{dn}{d\tau}\,,
\eea
which is a consequence of energy conservation in the fluid. Indeed, this equation can be derived through the thermodynamic laws of an adiabatically closed system. Similarly, the equation $P=0$ can be also obtained via thermodynamic relationships.

We further make the assumption that the thermal distributions of electrons and ions in the comoving frame of the fluid are approximated by the isotropic Maxwell-Jüttner distribution\cite{leung2011numerical}, which is characterized by a dimensionless temperature, $\T_j \equiv k_BT_j/m_j$, where $j\in\{\te, \ion\}$ indicates electron or ion, $k_B$ is the Boltzmann constant, $m_\te$ and $m_\ion$ are the masses  for  electron and ion, respectively. For ultra-relativistic particles with $\T_j \gg 1$, the internal enerty density becomes $\Xi_j\approx 3nm\T_j$. For non-relativistic particles with $\T_j \ll 1$, we obtain $\Xi_j \approx nm+ 3nm\T_j/2$. 

In this work, we would like to treat the electrons as ultra-relativistic particles at the horizon scale of the black hole, $\Xi_\te=3nm_\te\T_\te$, corresponding to $T_\te \gg 10^{9}$ K. The electrons far away from the black hole become mildly relativistic, which is not of interest to us \footnote{The term ``horizon scale" originates from the scale that can be resolved during the observation of EHT. It is approximately several times the size of $r_h$. According to Fig. \ref{nTEz}, for a Kerr black hole with $a=0.94$ in our model, within a range of at least $r \approx 6$, the electrons can be treated as ultra-relativistic.}. The hot plasma is most likely collisionless, resulting in a lack of thermal equilibrium between electrons and ions \cite{EventHorizonTelescope:2021srq}. For simplicity, we assume that the temperature ratio between ions and electrons at the horizon scale is characterized by a constant, given by $T_{\ion} =z T_\te$. Then, we have
\bea
\T_\ion=z\T_\te\frac{m_\te}{m_\ion}=\frac{z}{Z}\T_\te\,,
\eea
where $Z=\frac{m_\ion}{m_\te}\simeq1836$. The value of $z$ is uncertain and depends on the complex physical properties of the magnetofluid. To conduct a detailed analysis of the ion's temperature, we define the characteristic quantities to represent the orders of temperature at the horizon scale, denoted as $\T^\tc_\te = \T_\te(r_h) \gg 1 \,, \T^\tc_{\ion}  = \T_{\ion} (r_h)$, and $z_\tc=Z/\T^\tc_\te$. For example, in the case of M87$^\ast$, observations suggest that the electron temperature in the emitting region can reach $10^{11}$K \cite{EventHorizonTelescope:2019pgp, Vincent:2022fwj}. Therefore, we can estimate the characteristic quantities for M87$^\ast$ to be $\T_\te^\tc \simeq 15$, $ z_\tc \simeq 122$.

The properties of the two-component fluid studied in this work depend heavily on the nature of ions. For the ions, we will consider two limiting cases in order to obtain analytic results: either they are non-relativistic with $\T^\tc_\ion\ll1$, or they are ultra-relativistic with $\T^\tc_\ion\gg1$.  In the first scenario, we have $z\ll z_\tc$ and $\Xi_\ion \approx nm_\ion+ 3nm_\ion\T_\ion/2$, so that 
\bea
\Xi=\Xi_\te+\Xi_\ion=nm_\ion+3\left(\frac{1}{2}z+1\right)nk_BT_e\,.
\eea
For the non-relativistic case, the speed of sound is defined by
\bea
c_s =\sqrt{\frac{dp}{d\rho}}\,,
\eea
where $p$ and $\rho$ are the total pressure and the rest mass density of the fluid, respectively, \bea
p=nk_B\left(T_\te+T_\ion\right)=nk_BT_\te\left(1+z\right),\hspace{3ex}\rho=n\left(m_\te+m_\ion\right)=nm_\te(1+Z)\approx Znm_\te.\eea
 Thus, we can rewrite $\Xi$ in terms of $\rho$ and $p$,  
\bea
\Xi=\rho+\frac{3(z+2)}{2(z+1)}p\,.
\eea
From Eq. (\ref{teq}), we can read 
\bea\label{rp1}
\frac{dp}{d\rho}=\frac{5z+8}{3(z+2)}\frac{p}{\rho}=\frac{(5z+8)(1+z)}{3(z+2)(1+Z)}\T_\te\,,
\eea
and find 
\bea
c_s=\sqrt{\frac{(5z+8)(1+z)}{3(z+2)(1+Z)}\T_\te} \sim \sqrt{\frac{(5z+8)(1+z)}{3(z+2)z_c}}\,,
\eea
which is sub-relativistic as $z \ll z_\tc$. Moreover, Eq. (\ref{rp1}) can be equivalently rewritten as
\bea\label{TTT}
\f{d T_\te}{T_\te} = \f{2(1+z)}{3(2+z)} \f{d n}{n} \,,
\eea
which leads to 
\bea\label{T1}
T_\te(x^\mu)=\mT_0 [n(x^\mu)]^{\frac{2(1+z)}{3(2+z)}}\,,
\eea
with $\mT_0$ being the integration constant.

Next, we consider the second scenario in which the ions are ultra-relativistic with $\T_\ion\gg1$. In this case, both the electrons and the ions are ultra-relativistic particles, so that the rest mass density $\rho=nm_\te(1+Z)$ can be ignored compared to the pressure $p=3nm_\te\T_\te(1+z)$. As a result, we have
\bea
\Xi=3n\left(m_\te\T_\te+m_\ion\T_\ion\right)=3nm_\te\T_\te\left(1+z\right)=3p\,.
\eea
Note that in the relativistic case, the speed of sound is modified to
\bea
c_s=\sqrt{\frac{dp}{d\Xi}}\,,
\eea
so that we have $c_s=\frac{\sqrt{3}}{3}$ in the case that $\T_\ion\gg1$. In addition, similar to Eq. (\ref{TTT}), we obtain
\bea\label{TT}
\f{d T_\te}{T_\te} = \f{1}{3} \f{d n}{n} \,,
\eea
which leads to 
\bea\label{T2}
T_\te(x^\mu)=\mT_0 [n(x^\mu)]^{\frac{1}{3}}\,,
\eea
with $\mT_0$ being the integration constant.


\section{Analytical solutions}\label{sec3}
In this section, we present an analytical method to determine the properties of the magnetofluid. We have obtained the relation between the temperature $T_\te$ and number density $n$ in both the non-relativistic and ultra-relativistic limits, 
\bea\label{AT}
T_\te(x^\mu)=\left\{
\begin{aligned}
&\mT_0 [n(x^\mu)]^{\frac{2(1+z)}{3(2+z)}}\,, \quad &z\ll z_\tc\\
&\mT_0 [n(x^\mu)]^{\frac{1}{3}}\,. \quad &z\gg z_\tc\\
\end{aligned} \right.
\eea
The pressure is then obtained by $p = (1+z) n k_B T_\te$. Hence, the remaining physical quantities to be determined in the fluid are the number density $n$, the four-velocity $u^\mu$ and the magnetic field $B^\mu$ which are jointly governed by Eqs. (\ref{dum}), (\ref{neq}) and the ideal MHD condition, $F_{\m\n}u^\mu=0$.

As we will show shortly, the motion of particles in the above two limits can be treated approximately as geodesic motion. Then, the number density can be obtained from Eq. (\ref{neq}) by integrating along the geodesics. In general, it is not easy to derive the number density analytically from Eq. (\ref{neq}). However, if the left-hand side of Eq. (\ref{neq}) can be rearranged as a total derivative with respect to $\tau$, the solution for $n$ can be obtained directly. Although we do not know the general condition under which this case holds, we will provide a nontrivial example in Kerr spacetime in the following section. This example will be the main focus of this paper.

\subsection{The ballistic approximation}\label{ballistic}
In this subsection, we are going to deal with the relativistic Euler equations, Eq. (\ref{dum}). Let us first estimate  the magnitudes of the terms  on the right-hand side (RHS) of Eq. (\ref{dum}). We find that the coefficient of the first term behaves as
\bea
\frac{p}{\Xi+p}\simeq \left\{
\begin{aligned}
&\frac{2(1+z)}{2z_\tc+5z+8} \,, \quad &z\ll z_\tc\\
&\frac{1}{4}\,. \quad &z\gg z_\tc\\
\end{aligned} \right.
\eea
Hence, $\frac{p}{\Xi+p}$ is small in both limits, especially for $z\ll z_\tc$, so that the first term on the RHS of Eq. (\ref{dum}) can be neglected in the leading-order approximation. This approximation is consistent with the fact that the speed of sound discussed in Sec.\eqref{sec2} is sub-relativistic such that the gas pressure is not significant \cite{Weinberg:1972kfs}. For M87$^\ast$, $ z_\tc \simeq 122$, and as long as $z\ll 122$, the contribution of the gas pressure can be neglected.\footnote{Roughly speaking, the acceleration due to gravity near the horizon is estimated to be $a_g \sim GM/r_g^2 \sim c^2/r_g$, where $r_g = GM/c^2$ is the gravitational radius. The acceleration due to pressure gradients is $a_p \sim c_s^2/r_g$. Thus, for sub-relativistic speed of sound, $c_s \ll c$, the acceleration due to gravity is dominated, $a_g \gg a_p$.} This is consistent with the parameter range typically considered in numerical simulations of accretions of M87$^\ast$ \cite{EventHorizonTelescope:2019pgp}.

The second term on the RHS of Eq. (\ref{dum})  can be expanded as 
\bea
-\frac{\nabla_\m T^{\m\n}_\EM}{\Xi+p}=-\frac{\left[B^2u^\m\nabla_\m u^\n+B^2u^\n\nabla_\m u^\m+\left(u^\m u^\n+\frac{1}{2}g^{\m\n}\right)\nabla_\m B^2-B^\m\nabla_\m B^\n-B^\n\nabla_\m B^\m\right]}{\Xi+p}\,.
\eea
Consequently, the magnitudes of the coefficients in the second term can be approximated as
\bea
v_A^2=\frac{B^2}{\Xi+p}<\left\{
\begin{aligned}
&\frac{B^2}{\rho}\simeq\frac{B^2}{nm_\ion}\,, \quad &z\ll z_\tc\\
&\frac{B^2}{4p}\simeq\frac{ZB^2}{4nz\T_\te m_\ion}\,, \quad &z\gg z_\tc\\
\end{aligned} \right.
\eea
where $v_A$ is the Alf\'{v}en velocity, which has distinct expressions in both non-relativistic and ultra-relativistic limits. In this work, we assume that the magnetic field strength is dynamically unimportant such that $v_A\ll1$, and the second term on RHS of Eq. (\ref{dum}) can be neglected as well. Actually this assumption holds practical significance. The EHT collaboration estimated the number density for M87$^\ast$ using a spherical, one-zone toy model, yielding a range of $n =10^{10}\sim10^{11}\,\text{m}^{-3}$ \cite{EventHorizonTelescope:2019pgp}. However, a more realistic thick disk model developed in \cite{Vincent:2022fwj} estimated a maximum number density of $10^{12}\sim10^{13}\,\text{m}^{-3}$ to ensure an observed flux of $0.5$ Jy at 230 GHz. Regarding the associated magnetic field strength, the polarized images of M87$^\ast$ suggest parameter estimates ranging from $1$ to $30$ Gauss \cite{EventHorizonTelescope:2021srq}. Therefore, for our estimations, we take $B = 10$ Gauss and $n=5\times10^{12}\,\text{m}^{-3}$, yielding the following results
\bea
\left\{
\begin{aligned}
v_A&<0.1\ll1 \,, \quad &z\ll z_\tc\\
v_A&\ll0.05\ll1\,, \quad &z\gg z_\tc\\
\end{aligned} \right.
\eea
which indicates that our assumption on magnetic field and Alf\'{v}en velocity makes sense to the case of M87$^\ast$. Therefore, under our assumed conditions,  Eq. (\ref{dum}) can be simplified as follows:
\bea\label{ue}
u^\m\nabla_\m u^\n\simeq0\,,
\eea
which is the geodesic equation of the particle. The aforementioned approximation for fluid is known as the ballistic approximation, which has been employed to investigate various types of accretion streamlines \cite{Ulrich:1976zz,Tejeda:2012kb,Tejeda:2019lie}. However, the study of the thermodynamics and magnetic field structure of the ballistic magnetofluid at horizon scales of astrophysical black holes is lacking. It is also noteworthy to mention that the same treatment has been carried out in \cite{Mummery:2023tgh}, where they conducted a thorough examination of analytical thermodynamic solutions for intra-ISCO accretion in the equatorial plane.

Notably, the ballistic approximation is independent of the associated magnetic field, allowing us to solve for the four-velocity independently. Once we obtain the four-velocity, we can further determine the magnetic field through the ideal MHD condition $F^{\m\n}u_\m=0$. In this way, we can find  the solution for  physical parameters of fluid.

\subsection{Analytical solutions in Kerr spacetime}\label{theKerr}
In this subsection, we will analytically solve the fluid dynamics, namely the four-velocity, the particle number density, and the temperature of fluid in Kerr spacetime. The Kerr metric can be expressed in terms of the Boyer-Lindquist coordinates:
\bea
\mathrm{d}s^2&=&-\left(1-\frac{2Mr}{\Sigma}\right)\mathrm{d}t^2+\frac{\Sigma}{\Delta}\mathrm{d}r^2+\Sigma\mathrm{d}\theta^2+\left(r^2+a^2+\frac{2Mra^2}{\Sigma}\mathrm{sin}^2\theta\right)\mathrm{sin}^2\theta\,\mathrm{d}\phi^2-\frac{4Mra}{\Sigma}\mathrm{sin}^2\theta\,\mathrm{d}t\mathrm{d}\phi\, \nn\\
&=&g_{\mu\nu}\mathrm{d}x^\mu \mathrm{d}x^\nu.
\eea
The parameters $\Delta$ and $\Sigma$ are defined as
\bea
\Delta=r^2-2Mr+a^2\,,\quad \Sigma=r^2+a^2\cos^2\theta\,,
\eea
where $M$ and $a$ are the mass and the spin parameters of the Kerr black hole, respectively. For the sake of simplicity and without sacrificing generality, we shall henceforth adopt $M=1$. Then, the event horizon of the Kerr black hole resides at $r_h=1+\sqrt{1-a^2}$.

The timelike geodesic equations can be described in terms of three conserved quantities 
\bea
&&\Sigma u^t = \left[1 + \f{2  r \left(r^2+a^2\right)}{\D} \right] E - \f{2ar}{\D} L \label{t01}  \,, \\
&&\Sigma u^r =\s_r\sqrt{R}= \s_r\sqrt{\big[E(r^2+a^2)-aL \big]^2-\Delta\big[Q+(aE-L)^2+r^2\big]}\, , \label{R01} \\
&&\Sigma u^\t =\s_{\t}\sqrt{\T} = \s_{\t} \sqrt{Q-\cos^2{\t}\bigg[a^2(1-E^2)+\f{L^2}{\sin^2{\t}}\bigg]}\, ,\label{T01} \\
&&\Sigma u^{\phi} = \f{\D - a^2\sin^2{\t}}{\D \sin^2{\t}} L + \f{2ar}{\D} E \label{p01}  \, ,
\eea
where $E \ge 1, L$ are the energy and angular momentum per unit mass stemmed from the Killing vectors $\partial_t$ and $\partial_\phi$; $Q$ is the Carter constant per unit square mass, originating from the Killing 2-form of Kerr spacetime \cite{Carter:1968rr}; $\s_r, \s_{\t}$ denotes the sign of $u^r$ and $u^\t$, respectively.


Next, we proceed to study the stationary, axisymmetric streamlines of fluid in Kerr spacetime, particularly by calculating the expansion $\nabla_\m u^\m$ to solve the particle number conservation Eq. (\ref{neq}). Under the ballistic approximation, the streamlines become a bundle of geodesic described by Eqs. (\ref{t01})-(\ref{p01}). We consider that the streamlines extend within the spacetime outside an inner boundary shell denoted by $r_i \ge r_h$, with $\t_i$ representing the polar angle of the streamline on the shell. The streamlines with turning points in either the $r$ or $\t$ directions will not be considered here, as they are unnatural in a black hole accretion system \cite{Huerta:2006sc, Tejeda:2012kb}.

To evolve the streamlines of the fluid, we impose the distribution of conserved quantities of geodesics as initial conditions on the shell, $X = X(\t_i)$, where $X \in \{E,L,Q\}$. Using Eqs. (\ref{R01}), (\ref{T01}), for fixed $r_i$, the angle $\t_i$ can be written as a function of $r,\t$ and the conserved quantities along the geodesic, known as the inverse formula (Appendix \ref{AppB}). By substituting $X = X(\t_i)$ into the inverse formula, we can, in principle, solve for $\t_i = \t_i(r ,\t)$, depending on the specific form of $X(\t_i)$.  It should be emphasized that $ \t_i(r ,\t)$ must be single-valued, as two streamlines cannot intersect at a point. This condition is not always satisfied because we have not yet restricted the form of $X(\t_i)$. However, the specific form of $X(\t_i)$ that satisfies this condition is unknown,  and we assume that $\t_i$ is single-valued in the following derivation. Actually, the fluid we subsequently obtain indeed meets this condition.

The potentials $R, \T$ in Eqs. \eqref{R01}, \eqref{T01} can be rewritten as $R = R(r, X(\t_i)), \T = \T(\t, X(\t_i))$.  Since $\t_i$ is only a boundary value, $X(\t_i)$ is evidently invariant along the geodesic. This leads to
	\bea\label{aaa}
	\f{dR }{d\tau} = u^r \pl_r R \big|_{\t_i}  \, , \quad  \f{d \T}{d\tau} = u^{\t} \pl_{\t}  \T \big|_{\t_i} \,.
	\eea
	Then, we can calculate the expansion as
\begin{align}
\f{1}{\sqrt{|g|}} \pl_{\m}  (\sqrt{|g|}u^{\m}) &= \f{\s_r}{\S} \pl_r\sqrt{R}\big|_{\t} + \f{\s_\t}{\S \sin{\t}} \pl_\t \big(\sin{\t}\sqrt{\T}\, \big)\big|_{r} \nn \\
&=u^r\pl_r \log{\sqrt{R}}\big|_{\t} + u^\t \pl_\t \log{\sqrt{\T}}\big|_{r} + u^\t\partial_\t\log{\sin\t} \, ,
 \label{m1}
\end{align}
where we have used Eqs. (\ref{R01}), (\ref{T01}). Employing Eqs. (\ref{aaa}), (\ref{m1}), the expansion can be further expressed as
\begin{align} 
&u^r\pl_r \log{\sqrt{R}}\big|_{\t_i} + u^r\pl_{\t_i} \log{\sqrt{R}}\big|_{r} \pl_r\t_i + u^\t \pl_\t \log{\sqrt{\T}}\big|_{\t_i} +u^{\t} \pl_{\t_i} \log{\sqrt{\T}}\big|_{\t} \pl_{\t}\t_i + u^\t\partial_\t\log{\sin\t} \nn \\
&= \f{d}{d\tau} \big[\log{\sqrt{R \T}\sin{\t}}\,\big] + u^r\pl_{\t_i} \log{\sqrt{R}}\big|_{r} \pl_r\t_i + u^{\t} \pl_{\t_i} \log{\sqrt{\T}}\big|_{\t} \pl_{\t}\t_i \nn \\
&= \f{d}{d\tau} \big[\log{\sqrt{R \T}\sin{\t}}\,\big] +\pl_{\t_i} \big[\log{\sqrt{R^{-1}\T}}\, \big]\big|_{r,\t} u^{\t}\pl_{\t} \t_i  \, ,
 \label{m2}
\end{align}
where we have used $d\t_i /d\tau=(u^r\pl_r +u^{\t}\pl_{\t})\t_i =0$ to get the last term in the last line.
Here, we have taken into account of the effect of variation of the conserved quantities with respect to the boundary value $\t_i$.
We notice that if the second term on the RHS of Eq.~\eqref{m2} can be neglected, the expansion becomes a total derivative term with respect to $\tau$. This occurs if the streamlines satisfy one of the following conditions: $(1)$  The variations of the conserved quantities with respect to $\t_i$ are tiny on the boundary shell, then $\pl_{\t_i} R |_{r} = \f{dX}{d\t_i} \pl_{X}R|_{r}  \approx 0, \pl_{\t_i} \T|_{\t}  = \f{dX}{d\t_i} \pl_{X}\T|_{\t}  \approx 0$; $(2)$ $u^{\t}$ is directly equal to zero. We assume that the fluid model we are considering satisfies at least one of the above conditions. Thus, with Eqs. (\ref{neq}), (\ref{AT}), we obtain the explicit expressions for the number density and the temperature 
\bea\label{nT1}
n(r,\t) = n(r_i,\t_i) \sqrt{\f{R(r_i)\T(\t_i)}{R(r)\T(\t)}}\f{\sin{\t_i}}{\sin{\t}} \,,
\eea
and 
\bea\label{nT2}
T_\te(r,\t)= \left\{
\begin{aligned}
&T(r_i,\t_i) \left[\f{R(r_i)\T(\t_i)}{R(r)\T(\t)}\right]^{\f{(1+z)}{3(2+z)}}\left[\f{\sin{\t_i}}{\sin{\t}}\right]^{\f{2(1+z)}{3(2+z)}}\,, \quad &z\ll z_\tc\\
&T(r_i,\t_i) \left[\f{R(r_i)\T(\t_i)}{R(r)\T(\t)}\right]^{\f{1}{6}}\left[\f{\sin{\t_i}}{\sin{\t}}\right]^{\f{1}{3}}\,, \quad &z\gg z_\tc\\
\end{aligned} \right.
\eea
where $n(r_i, \t_i), T(r_i, \t_i)$ are introduced as the boundary values, with  $\t_i$ being determined by the inverse formula $\t_i(r,\t)$. At this point, we have obtained the explicit solutions for the thermodynamics of the fluid, i.e., the particle number density $n(r, \t)$ (Eq. (\ref{nT1})), the temperature $T_\te(r, \t)$ (Eq. (\ref{nT2})). As mentioned before, the expressions are inapplicable at the turning points in radial or angular directions, where Eqs. (\ref{nT1}), (\ref{nT2}) have poles.

Different from the torus model \cite{abramowicz1978relativistic,Komissarov:2006nz} with pure toroidal fluid, the model we are studying describes a fluid configuration with nonzero poloidal velocity. In the near-horizon region of an astrophysical black hole, the poloidal flow is a more realistic scenario, where gravity plays a dominant role, and heat transfer primarily occurs through advection \cite{Narayan:1994is, Narayan:1994et}.
The fluid configurations satisfying $u^{\t} = 0$ are similar to simulated accretion flows and FWJs at horizon scales. Hence, in the following discussion, we will focus on this scenario, providing an explicit solution and conducting an analysis accordingly.


\subsection{The conical solution}\label{Con}

In this subsection we delve into the fluid satisfying the condition $u^\t=0$. In this case,  the streamline maintains a constant value of $\t$ in the polar direction. This type of motion can be  achieved when $\T=\partial_\t\T=0$, resulting in the following expression:
\bea\label{LQ}
L = \pm_L a\sqrt{E^2-1}\sin^2{\t} \,\, , \quad  Q = -a^2(E^2-1)\cos^4{\t}
\eea 
with ``$\pm_L$'' denoting the sign of $L$. Moreover, one requires $\pl^2_{\t}\T = -8a^2(E^2-1)\cos^2{\t} \leq 0$ in order to have stable  geodesics. As the fluid is foliated by the streamlines on conical surfaces, we call the solution satisfying Eq. (\ref{LQ}) the conical solution. The radial potential can be expressed as 
\bea\label{Rc}
R_{\text{c}}(r,\t)&=& (E^2-1)r^4 + 2r^3 + 2a^2(E^2-1)\cos^2{\t}r^2  \nn \\
&+& 2a^2\bigg[\big(E\mp_L \sqrt{E^2-1}\sin^2{\t}\big)^2-(E^2-1)\cos^4{\t}\bigg]r + a^4(E^2-1)\cos^4{\t}\, ,
\eea
which is non-negative as long as $E \geq 1$. The 4-velocity is now of the form 
\bea\label{u}
&&u^t = E\bigg[1+\f{2r(r^2+a^2)}{\D\S}\bigg] \mp_L \sqrt{E^2-1}\f{2a^2r\sin^2{\t}}{\D\S} \, , \nn \\
&&u^r = \s_r \f{\sqrt{R_{\text{c}}}}{\Sigma} 
\, , \quad u^{\t} = 0\, , \nn \\
&&u^{\p} = E\f{2ar}{\D\S} \pm_L  \sqrt{E^2-1}\f{a(\D-a^2\sin^2{\t})}{\D\S} \, .
\eea
The number density and temperature get simplified to
\bea\label{nT11}
n(r,\t) = n(r_i,\t) \sqrt{\f{R_{\text{c}}(r_i,\t)}{R_{\text{c}}(r,\t)}} \,,
\eea
and 
\bea\label{nT22}
T_\te(r,\t)= \left\{
\begin{aligned}
	&T(r_i,\t) \bigg[\f{R_{\text{c}}(r_i,\t)}{R_{\text{c}}(r,\t)}\bigg]^{\f{(1+z)}{3(2+z)}}\,, \quad &z\ll z_\tc\\
	&T(r_i,\t) \bigg[\f{R_{\text{c}}(r_i,\t)}{R_{\text{c}}(r,\t)}\bigg]^{\f{1}{6}}\,, \quad &z\gg z_\tc\\
\end{aligned} \right.
\eea
where $n(r_i,\t)$ and $T(r_i,\t)$ are the boundary values of $n$ and $T_\te$, respectively. For convenience in the following studies, we select a Gaussian distribution in the $\t$ direction for $n(r_i, \t)$, and let $T(r_i, \t)$ be a constant for the conical solution:
\bea\label{02}
n(r_i,\t)=n_i \exp \bigg[ -\bigg(\f{\sin{\t}-\sin{\t_J}}{\s}\bigg)^2 \bigg] \, , \quad T(r_i,\t) =T_i \, ,
\eea
where $\t_J$ is the mean position in $\t$ direction, and $\s$ describes the standard deviation of the distribution. As mentioned before, the inner boundary $r_i$ can be positioned at any location outside the horizon. For the sake of convenience, let us choose to place it at the horizon in subsequent studies, that is $r_i=r_h$.

\subsection{The magnetic field}\label{magnetic}

Next, let us proceed to address the magnetic field structure accompanying the axisymmetric and stationary magnetofluid. The non-vanishing components of $F_{\m\n}$ are described by the gauge potential $A_\m(r, \theta)$ \cite{Ruffini:1975ne},
\bea\label{RfF}
&& F_{r\p} = \pl_r A_{\p} \, , \quad  F_{\t\p} = \pl_\t A_\p \, ,\quad F_{r\t} =\partial_r A_\t-\partial_\t A_r \, .
\eea 
By considering the Maxwell equation $\nabla_\mu \,^\ast F^{\mu\p}=0$, and the ideal MHD condition, $F^{\p\mu}u_\mu=0$, one finds 
\bea\label{uA}
(u^r \pl_r  + u^\t \pl_\t )A_\p =0\,,
\eea
which means that $A_\p$ is invariant along the streamline. This indicates that the general form of $A_\p$ is $A_\p=A_\p(\theta_i)$, where $\t_i = \t_i(r,\t)$ is the polar angle on the boundary shell. Thus, one can solve the field configuration if the streamlines are known. Moreover, combining Eq. (\ref{uA}) and the $t$ component of the Maxwell equation $\nabla_\mu \,^\ast F^{\mu t}=0$ gives 
\bea\label{RfFF}
F_{r\t} = \f{u^\p}{u^r} \pl_\t A_\p\,.
\eea
With the expressions of $F_{\m\n}$, the magnetic field $B^{\mu} = -\,^\ast F^{\mu\nu} u_{\nu}$ can be obtained accordingly, that is,
\bea\label{B4}
&&B^t = \f{1}{\sqrt{|g|}} \f{\pl_{\t}A_{\p}}{u^r} \big( u^t u_t +1 \big) \, , \nn \\
&&B^i = \f{1}{\sqrt{|g|}} \f{\pl_{\t}A_{\p}}{u^r} u_t u^i \, , \quad  i = r,\t,\p \, .
\eea
The spatial component $B^i$ is parallel to $u^i$, indicating that the magnetic field is frozen into the streamlines. It is important to emphasize that the aforementioned derivation (Eq.\eqref{RfF}-Eq.\eqref{B4}) solely relies on the assumption of stationarity, axisymmetry, and the ideal MHD condition \cite{Ruffini:1975ne}, without utilizing explicit metric and the ballistic approximation. 

For the sake of further discussion about the magnetic field in the conical solution in Sec. \ref{Con}, we derived the magnetic field measured in the comoving frame of the fluid in the case of $u^\t = 0$. The setup of tetrads can be found in Appendix \ref{AppC}. In the comoving frame, the magnetic field components take
\bea\label{BFRF}
B^{(0)} = B^{(2)} = 0 \, , \quad B^{(1)} = \f{ \pl_\t A_\p}{\hat{u}\sqrt{|g|}}  \sqrt{g_{rr}g_{\p\p}}\, \o^\p \, , \quad  B^{(3)}  =  -\a  \f{ \pl_\t A_\p}{u^r\hat{u}\sqrt{|g|}}  (u^tu_t +1) \,,
\eea 
where $\a$ is the lapse function, $\o^\p$ is the angular velocity of frame dragging, and $\hat{u} = \sqrt{(u^{(r)})^2+(u^{(\p)})^2} = \sqrt{-1 + (\a u^t)^2}$ is the magnitude of the fluid velocity in zero-angular momentum observers (ZAMOs). It can be seen that $B^{(1)} $ is purely induced by the frame dragging. Besides, Eq.~\eqref{uA} shows that $\Psi \equiv \pl_\t A_\p$ is a function of $\t$ for conical solutions. For simplicity, we employ the split monopole configuration, which is the simplest model for the global magnetic field configuration \cite{Blandford:1977ds},
\be\label{Psi0}
\Psi = \Psi_0 \, \text{sign}(\cos{\t})  \sin{\t}  \, ,
\ee
where $\Psi_0$ is a constant. The introdution of $\sin{\t}$ is to ensure regularity at the poles $\sin{\t} = 0$. The sign function ensures that the black hole has zero magnetic charge. Actually, numerical studies indicate that a split monopole configuration naturally emerges in the near-horizon region with a initial uniform magnetic field \cite{Komissarov:2004qu, Komissarov:2004ms}.

It should be emphasized that although the magnetic field is dynamically unimportant under the ballistic approximation, it can still be constrained by the ideal MHD conditions to a global factor along the streamlines. For the magnetized torus model \cite{Komissarov:2006nz}, the magnetic pressure affects the torus structure, but additional simplifying assumptions are also required to obtain explicit expressions.

We must acknowledge that the assumption of a dynamically unimportant magnetic field is an oversimplification when dealing with conventional magnetofluids. More accurately, the magnetic field structure is coupled to the plasma fluid in MHD. To accurately determine the magnetic field configuration, it is necessary to solve both the Euler equations and the divergence-free Maxwell equation.

\section{Applications to accretion and jet flow}\label{sec4}
In this section, we apply the conical solution discussed in the previous section to the study of accretion flow and jet. In the case of accretion flow, we manage to establish a thick accretion flow model at the horizon scale. As for the jet flow, we present a FWJ model. These two models have the advantage of being analytically solvable.

\subsection{Accretion flow}\label{accd}

It is well known that the standard accretion disk model describes a geometrically thin and cold accretion flow in the equatorial plane \cite{page1974disk}, where viscosity and radiation play significant roles. However, observations indicate the existence of diverse types of thick disk morphologies around active galactic nuclei \cite{EventHorizonTelescope:2019pgp,EventHorizonTelescope:2022urf}. For instance, in the case of M87$^\ast$, studies suggest the presence of a Radiatively Inefficient Accretion Flow (RIAF) surrounding it \cite{EventHorizonTelescope:2019pgp}. Furthermore, when approaching the event horizon, the behaviors of the flows are primarily governed by gravity. With the development of the EHT imaging, the understanding of the plasma physics in this region becomes increasingly important. 
Motivated by this fact, we will utilize the analytical model developed in the last section to investigate the thick accretion flow close to the black hole. With the explicit expressions for $n,\, T$, and $B^\m$, the model offers a direct avenue to study black hole imaging, which will be demonstrated in \cite{part2}. 

In this subsection, we discuss the thermodynamics and magnetic field structure of accretion flows exhibiting conical motions. We consider a thick accretion disk that exhibits equatorial symmetry, described by $\t_J=\pi/2,\s=1/5$.  However, it is generally uncertain to determine the boundary conditions for such accretion flow. In the case of an idealized thin disk, the accretion flow near the horizon falls in from the ISCO, and its boundary values are taken at the ISCO, while for a thick disk, such a boundary does not exist, and the large-scale dynamics are also unknown. Here we simply consider a freely falling flow that satisfies
\bea
\sigma_r = -1 \, , \quad E = 1 \,.
\eea
In this case, the radial potential simplifies to $R_{\text{c}}(r)=2r(r^2+a^2)$. From Eqs. (\ref{nT11}) and (\ref{nT22}), the thermodynamics for the free falling accretion is characterized by
\bea\label{nacc}
n(r,\t) = n_i \exp \left[-25(\sin{\t}-1)^2\right] \sqrt{\f{2r_h^2}{r(r^2+a^2)}} \,,
\eea
and
\bea\label{Tacc}
T_\te(r)= \left\{
\begin{aligned}
&T_i \left[\sqrt{\f{2r_h^2}{r(r^2+a^2)}}\right]^{\f{(1+z)}{3(2+z)}}\,, \quad &z\ll z_\tc\\
&T_i \left[\sqrt{\f{2r_h^2}{r(r^2+a^2)}}\right]^{\f{1}{6}}\,, \quad &z\gg z_\tc\\ 
\end{aligned} \right.
\eea
Note that we have set $r_i = r_h$ when deriving the above equations. Compared to semi-analytical models \cite{Broderick:2008qf, Pu:2016qak, Pu:2018ute}, the dependence of the number density and temperature on $r$ in our model is non-power-law in general. However, far from the horizon, we still have $n \propto r^{-1.5}$, $T \propto r^{-\f{(1+z)}{2(2+z)}}$ for non-relativistic ions and $T \propto r^{-0.25}$ for relativistic ions. In contrast, the number density distribution in the $\t$ direction is determined by the choice of boundary condition, for which we  employ a Gaussian distribution. In this work, we have only presented the accretion flow characterized by the conical motion. It is worth considering more general streamline configurations in practice. Nonetheless, the conical model roughly captures the structure of time-averaged accretion flows near the event horizon generated by  GRMHD simulation \cite{Shen:2023nij}.

\subsection{Funnel wall jet}

Astrophysical black holes with high spin often exhibit relativistic jets about their rotational axis while accreting matter. The observed jet in M87* is well collimated at distance, which means it has a parabolic shape in the jet acceleration zone \cite{Lu:2023bbn}. In our model, by choosing proper conserved quantities in Eq.~\eqref{nT1}, we can in principle construct a FWJ of parabolic shape. However, near the horizon, the jet shape might be approximated as a conical shape, as long as it is generated by the B-Z process \cite{Blandford:1977ds} instead of the B-P process \cite{Blandford:1982di}. For simplicity, we would like to employ the conical solution to describe the FWJ in the following.

We designate the angle of the FWJ's center to be $\t_J=\pi/4$ and assume $\sigma=1/10$. Furthermore, since in the case of a jet flow, the kinetic energy of particles at infinity can be non-zero,  we can consider taking $E\ge1$ in this scenario. Note that while $L$ and $Q$ are determined by Eq.~\eqref{LQ}, the dependence of $E$ on $\t$ remains arbitrary. For simplicity, we will treat it as a constant in this study. Moreover, numerical studies indicate the presence of inward FWJs at horizon scales \cite{Globus:2013fla,Globus:2014eaa}. Thus, the radial velocity $u^r$ can be either outward or inward, which means both positive and negative values of $\sigma_r$ are permissible. However, from Eqs. (\ref{nT1}), (\ref{nT2}), and (\ref{B4}), it can be observed that the sign of $\sigma_r$ does not affect the particle number density $n$ and the electron temperature $T_\te$, and it only alters the sign of certain components of the magnetic field.  Hence, we may consider $\sigma_r=+1$ for the jet flow.

We want to emphasize that our model is not able to fully characterize the jet dynamic, as we have neglected the magnetic pressure and viscosity in the funnel region, which might be significant for jet acceleration and collimation. Nevertheless, the conical FWJ model is quite simple and has the potential to capture the morphology of the horizon-scale FWJ, thus contributing to the study of black hole imaging. Specifically, the direct and lensed images of the double-cone structure might be of interest; the Doppler shift caused by the inward and outward flows significantly affects the imaging process as well.

\subsection{Figures}\label{Fs}

To gain a more intuitive understanding, we present figures illustrating the particle number density $n$, electron temperature $T_\te$, and magnetic field $B^\m$. For convenience, we introduce Cartesian coordinates as
\bea
\mathcal{X}=r \sin\theta \cos\phi\,,\quad \mathcal{Y}=r\sin\theta\sin\phi\,,\quad \mathcal{Z}=r \cos\theta\,.
\eea

\begin{figure}[ht!]
\centering
\includegraphics[width=5.5in]{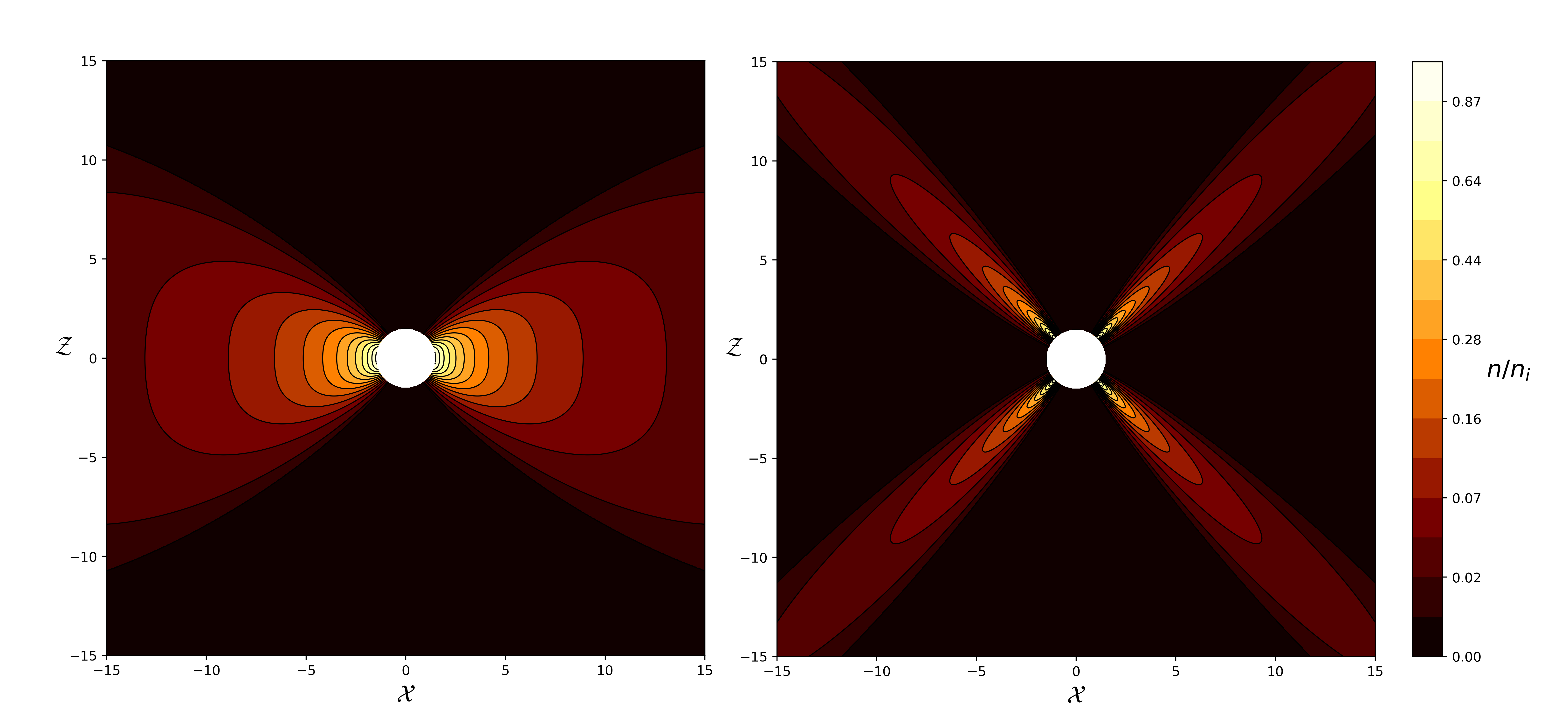}
\centering
\caption{The particle number density in the $\mathcal{X}$-$\mathcal{Z}$ plane. The left figure corresponds to the accretion disk, for which we select $E=1, \sigma=1/5$ and $\theta_J=\pi/2$, while the right one corresponds to the FWJ, for which we take $E=1, \sigma=1/10$ and $\theta_J=\pi/4$. The unit of length is the gravitational radius $GM/c^2$.}
\label{nden}
\end{figure}

In Fig. \ref{nden}, we illustrate the two-dimensional density plot of the particle number density $n/n_i$ in the $\mathcal{X}$-$\mathcal{Z}$ plane. The left panel displays the results for the accretion disk, while the right panel shows the results for the FWJ. The white region at the center of each figure represents the black hole. Here and throughout, we assume a black hole spin parameter of $a=0.94$. 

\begin{figure}[ht!]
\centering
\includegraphics[width=6.8in]{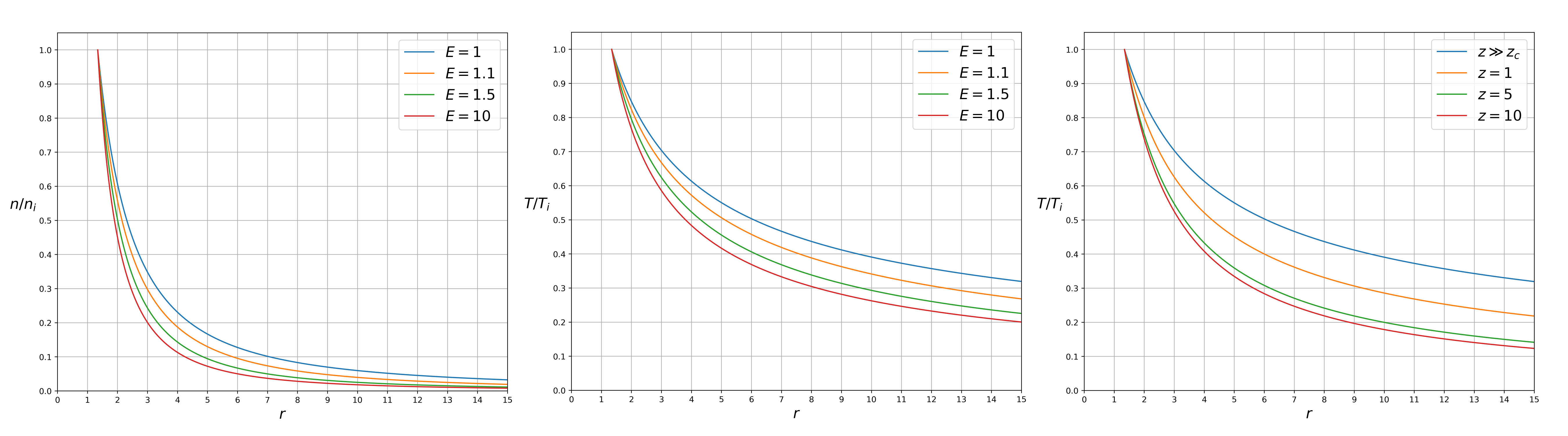}
\centering
\caption{Particle number densities $n$ and electron temperatures $T_\te$ as functions of $r$. In the preceding two plots, we have respectively chosen $E$ values of $1, 1.1, 1.5$, and $10$, and $z\gg z_\tc$ for the temperature. In the third plot, we have taken $z=1, 5, 10$, $z\gg z_\tc$, and kept $E = 1$. The unit of length is the gravitational radius $GM/c^2$.}
\label{nTEz}
\end{figure}

Upon examining Eqs. (\ref{nT11}), (\ref{nT22}), it becomes apparent that the dependence of particle number density and electron temperature on $r$ and $\t$ is not decoupled, as evidenced by the expression for $R_\tc$ (Eq. (\ref{Rc})), unless $E = 1$. However, since the conical motion occurs solely along the $r$ direction, it is meaningful to study the variations of particle number density and temperature separately with respect to $r$. In the cases of the accretion disk and FWJ, the differences lie in the polar positions and the assigned values of $E$: for the accretion disk, $E=1$, while for the FWJ, $E\ge1$ can take on various values. Additionally, as the ratio $z$ between ion temperature and electron temperature changes, the behavior of the electron temperature will also undergo corresponding variations.

In Fig. \ref{nTEz}, we present the analytical results depicting the variations of particle number density and temperature as a function of $r$. To focus solely on the effects of $E$ and $z$, we set $\t = \t_J$ in Eq. (\ref{02}) in all cases to eliminate the influence of the Gaussian distribution. Moreover, we have set $z_\tc = 122$ to match the astronomical environment of M87$^\ast$. In the left and middle plots, we have selected $E=1, 1.1, 1.5, 10$ respectively. In the right plot, we have modified the value of $z$ to be $1, 5, 10$, and significantly greater than $122$, while keeping $E =1$. From these plots, we can observe that both $n$ and $T_\te$ are decreasing functions of $r$. Furthermore, as $E$ increases, the decreasing rate becomes larger. Additionally, concerning the electron temperature, when the ions are non-relativistic with $z\ll 122$, an increasing $z$ leads to a faster  decreasing in $T_\te$. However, in the case of ultra-relativistic ions with $z\gg 122$, the decreasing rate   is actually the smallest. Comparing the quantities of $n$ and $T_\te$ under the same conditions, it is evident that the decay rate of $n$ is larger.

\begin{figure}[ht!]
\centering
\includegraphics[width=6.8in]{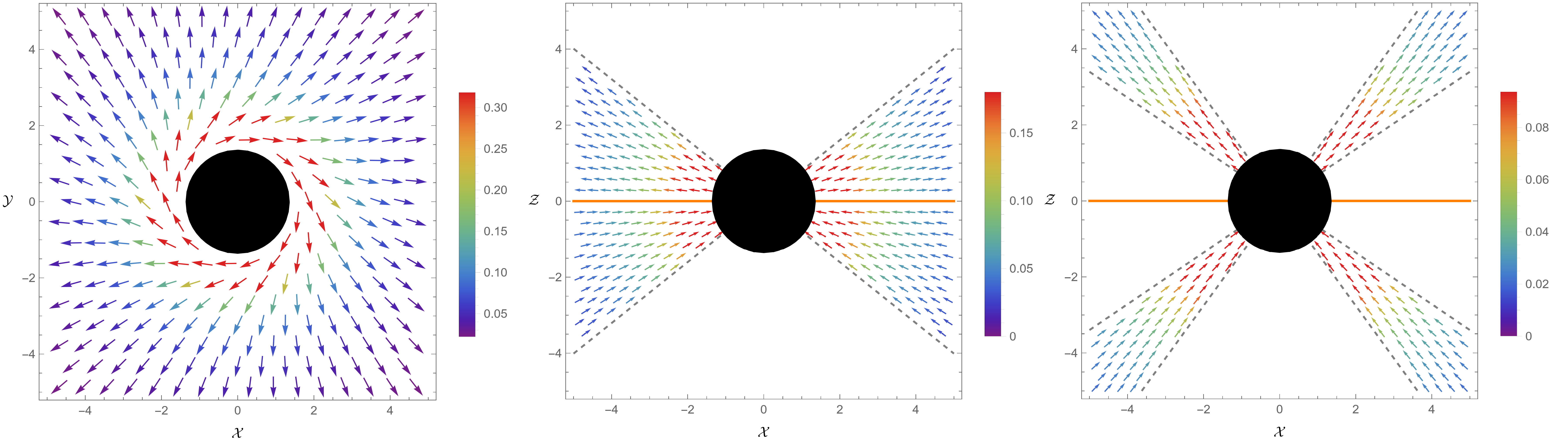}
\centering
\caption{Magnetic field configurations in the accretion disk and the FWJ models. The left plot shows the magnetic field in the equatorial plane in the accretion disk model, the middle and the right ones depict the poloidal field structures in the $\mathcal{X}$-$\mathcal{Z}$ plane for the accretion disk and FWJ models, respectively. In all the plots, we set $E=1$. In the case of the FWJ model, while the value of $E$ can be adjusted, it only affects the magnitude of the magnetic field and not its orientation. The unit of length is the gravitational radius $GM/c^2$.}
\label{Bp}
\end{figure}

\begin{figure}[ht!]
\centering
\includegraphics[width=6.8in]{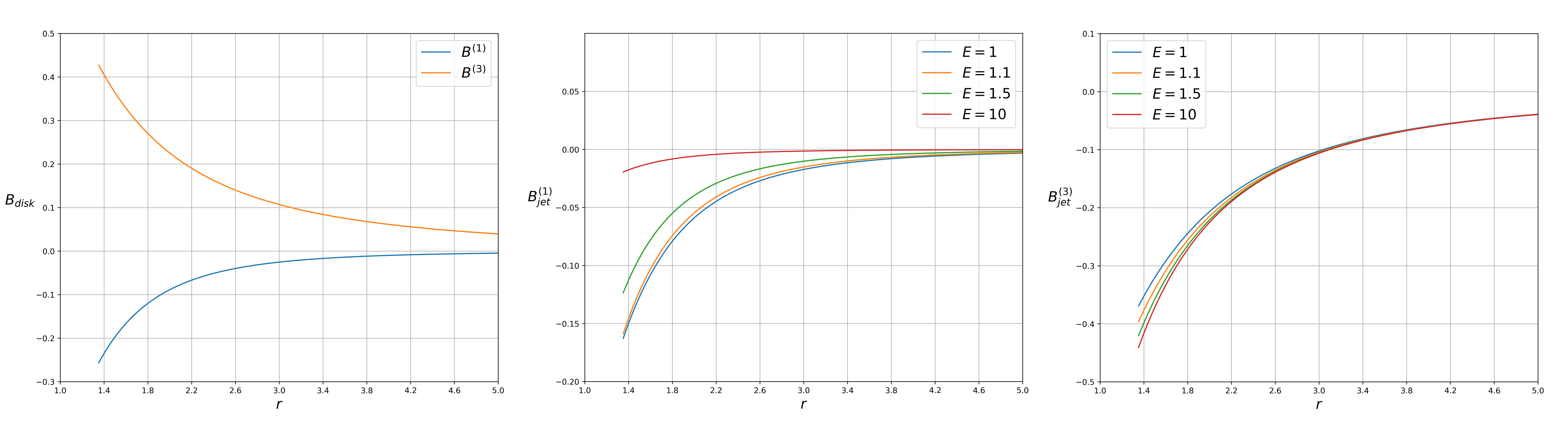}
\centering
\caption{The variation of the magnetic field with respect to $r$ in the comoving frame of the fluid. The left plot illustrates the variations in the accretion disk model, with $E=1$. The subsequent two diagrams depict the outcomes of the FWJ model, with the choice $E=1, 1.1, 1.5, 10$ respectively. The unit of length is the gravitational radius $GM/c^2$.}
\label{Br}
\end{figure}

Now let us shift our attention to the magnetic field structures. For simplicity, we choose $\Psi_0=-1$ in the subsequent discussion. Fig. \ref{Bp} displays the magnetic fields for both the thick disk and FWJ. The first panel illustrates the magnetic field configuration in the equatorial plane for the disk model\footnote{Indeed, according to Eq. (\ref{Psi0}), it can be observed that $\Psi=0$ in the equatorial plane, indicating the absence of magnetic field distribution. However, we can consider a slight deviation of $\t$ from $\pi/2$, which can be approximated as the equatorial plane. In \cite{Ruffini:1975ne}, the corresponding discussions also utilize the expression of magnetic field in the equatorial plane. In Fig. \ref{Bp}, we have chosen $\t=\pi/2-0.01$.}. The middle and right panels depict the magnetic fields in the $\mathcal{X}$-$\mathcal{Z}$ plane for the disk and FWJ models, respectively. The left plot clearly reveals that in the near-horizon region, the magnetic field takes on a spiral shape. As $r \rightarrow r_h$ one finds $B^r/B^{\p} \rightarrow 0$ with Eq.~\eqref{B4} and Eq.~\eqref{u}, indicating that due to frame dragging, the magnetic field exhibits an extremely spiral pattern when $r \rightarrow r_h$. Therefore, the polarized images of the model are expected to carry the information about the magnetic field structure, fluid streamlines, and black hole spin. Once moving away from the black hole, the magnetic field direction becomes predominantly radial. In the $\mathcal{X}$-$\mathcal{Z}$ plane, the magnetic field structure is similar to the disk model and the FWJ model. It follows a primarily radial pattern, but with a distinction: in the northern hemisphere, the magnetic field direction diverges away from the black hole; while in the southern hemisphere, the magnetic field points towards the black hole.

In Fig. \ref{Br}, we present the variation of the magnetic field with respect to $r$ in the comoving frame of the fluid. By examining the left plot, which shows the changes of the magnetic field in the disk model, we observe that $B^{(1)}<0$ and $B^{(3)}>0$, but both magnitudes decrease as the radius increases. The middle and right plots in Fig. \ref{Br} show the variations of magnetic fields in the FWJ model. We find that both $B^{(1)}$ and $B^{(3)}$ within the jet are negative, and their magnitudes exhibit a monotonically decreasing trend. Moreover, for all the parameters chosen, $|B^{(3)}|$ is much larger than $|B^{(1)}|$ throughout the streamlines, signifying that the magnetic field in the comoving frame of the fluid is primarily aligned along the direction of flow. It is evident that the spatial arrangement of the magnetic field within the FWJ exhibits a noticeable difference compared to the one within the thick disk, despite their analogous mathematical representations.

\section{Summary and discussions}\label{sec5}

In this work, we have developed a simplified magnetofluid model that enables us to study analytically the kinematics and thermodynamics of the magnetofluid surrounding rotating black holes. For the fluid part, we took it to be an ideal, adiabatic plasma consisting of two components, electrons and ions, having different temperatures but being characterized by a constant temperature ratio $z$. The horizon-scale electrons are treated as ultra-relativistic thermal particles with $T_\te \gg 10^{9}$K, corresponding to the hot plasma around a supermassive black hole. For ion temperature, there exists a characteristic quantity, denoted as $z_\tc$, such that the ions with $z \ll z_\tc$ can be considered non-relativistic, while the ions with $z \gg z_\tc$ are considered ultra-relativistic. In our analysis, we have focused solely on these two limiting cases, for the sake of simplicity. As for the magnetic field, we assume that the magnetic pressure is weak, resulting in the Alf\'{v}en velocity being sub-relativistic in the plasma.

Based on the aforementioned treatments, the flows are primarily governed by gravity, allowing us to employ the ballistic approximation for the magnetofluid. We have analyzed the relevant parameter space and justified the applicability of the ballistic approximation to the magnetofluid surrounding M87$^\ast$. Subsequently, by examining the stationary, axisymmetric, and ballistic fluid configuration in Kerr spacetime, we  discovered that the thermodynamics can be solved analytically (Eqs. (\ref{nT1}), (\ref{nT2})) for the fluid characterized by a slow variation of conserved quantities between neighboring streamlines or the fluid satisfying $u^{\t} = 0$, the latter being referred to as the conical solution. In particular, since the conical solution exhibits features similar to the simulated accretion flow and FWJ close to the black hole, we focused on this scenario and provided detailed discussions in Sec. \ref{Con}, including the structure of the accompanying magnetic field in Sec. \ref{magnetic}. Furthermore, we have explored the potential applications of our model to real astronomical environments and utilized the conical solution to describe thick accretion disks and FWJs at the horizon scale, as shown in Sec. \ref{sec4}. Additionally, we have presented graphs in Sec. \ref{Fs} to facilitate a more intuitive understanding of the characteristics of the thick disk and FWJ described by the conical solution.

Compared to previous studies, our model presents a novel magnetofluid configuration with inward or outward poloidal flows that extend beyond the equatorial plane, aiming to capture the fluid streamlines and thermodynamics at the horizon scale. The presence of geometrically thick, optically thin magnetofluid with poloidal magnetic field close to M87$^\ast$ \cite{EventHorizonTelescope:2019dse, EventHorizonTelescope:2021srq} provides strong motivation for the development of our model. Within the framework of the ideal magnetofluid, we have employed the ballistic approximation to facilitate analytical discussions. However, there are a few subtle issues  that need to be addressed. Firstly, unlike the studies that used a similar approximation to study the formations of thin disks by directing the streamlines into the equatorial plane \cite{huerta2007simple, Tejeda:2012kb}, we treated the horizon-scale plasma as a ballistic fluid to solve its thermodynamics. As our focus was solely on the physics at the horizon scale, we chose the boundary values at the horizon. Therefore, we need further physical considerations or inputs from the observation to determine the appropriate boundary conditions. For instance, if the EHT observations can successfully capture the inner shadow of the horizon, the intensity distribution of its edge contour undoubtedly has the potential to reflect specific boundary conditions. Secondly, the studies of the EHT collaboration suggest that the accretion of M87* is mostly in a magnetically arrested disk (MAD) state \cite{EventHorizonTelescope:2021srq}, and the global magnetic field is dynamically important to generate the jet. Our model, however, is not suited for strongly magnetized flows, such as the near-axis jet, where the magnetic field utterly dominates the plasma. This type of plasma can be described by force-free electrodynamics \cite{Gralla:2014yja}. Nonetheless, within the disk region, the fluid description remains valid, and the time-averaged accretion structure is primarily determined by the fluid part. Thus, our model remains applicable to the time-averaged accretion flow at the horizon scale, where gravitational acceleration dominates.

Despite these, further studies can be conducted with our simplified model. With the explicit expressions for $n, T_\te$, and $B^\m$, our model provides a direct avenue for studying black hole imaging. Specifically, for the FWJ model, the gravitational lensing of the double-cone structure may be of interest, and the observed intensities will be affected by the Doppler shift caused by the inward and outward flows. In our upcoming work \cite{part2}, we will demonstrate the imaging features of the thermal synchrotron radiation from thick disks and FWJs using the conical solution in this work.  Besides, it is shown in Fig.~\ref{Bp} that the magnetic field in Eq. (\ref{B4}) exhibits a highly spiraling nature near the horizon, which influences both the anisotropic emissivity and plasma polarization. It may lead to the polarization patterns of specific features, which could be checked in the images of the black holes. 

\section*{Acknowledgments}
We thank Zhong-Ying Fan and Ye Shen for helpful discussions. The work is partly supported by NSFC Grant No. 12275004, 12205013 and 11873044. MG is also endorsed by ”the Fundamental Research Funds for the Central Universities” with Grant No. 2021NTST13.

\appendix			 

\section{Inverse formula for $\t_i$}\label{AppB}

The expression for $\t_i(r, \t)$ can be obtained by solving the timelike geodesic equations in Kerr spacetime, resulting in the so-called inverse formula expressed in terms of elliptic functions. For $Q > 0$, the expression takes the following form
\bea
\cos{\t_i} &=&  \sqrt{u_+} \, \text{sn} \bigg( F_{\t} + \s_{\t} a\sqrt{-u_-}\, I_r  \, \bigg| \, \f{u_+}{u_-} \bigg) \, , \nn \\
\mbox{with}~~F_{\t} &=& F\bigg( \arcsin\bigg(\f{\cos{\t}}{\sqrt{u_+}}\bigg) \, \bigg| \, \f{u_+}{u_-} \bigg)  \, .
\eea
For $Q < 0$, the expression takes the form
\bea
\cos{\t_i} &=& h_{\t} \sqrt{u_-} \, \text{dn} \bigg( F_{\t} + h_{\t}\s_{\t} a\sqrt{u_-}\, I_r   \, \bigg| \, 1- \f{u_+}{u_-}  \bigg) \, , \nn \\
F_{\t} &=& F\bigg( \arcsin\sqrt{\f{\cos^2{\t}-u_-}{u_+-u_-}} \, \bigg| \, 1-\f{u_+}{u_-} \bigg) \, ,
\eea
where $h_{\t} = \text{sign}(\cos{\t})$, $u_{\pm} = \D_{\t} \pm \sqrt{\D^2_\t + Q(E^2-1)^{-1}a^{-2}}$ with $\D_\t = \f{1}{2}(1- (Q+L^2)(E^2-1)^{-1} a^{-2})$.
The radial integral takes
\bea
I_r = \sqrt{E^2-1} \int^{r}_{r_i} \f{dr}{ \sqrt{R}} \, .
\eea
The inverse formula for null geodesics is identical to that for timelike geodesics, with the conserved quantities are replaced by impact parameters $L/\sqrt{E^2-1} \rightarrow \lambda$, $ Q/(E^2-1)\rightarrow \eta$. For the case of freely falling flow with $E = 1, L = 0, Q \ge 0$, we have $R = 2r(r^2+a^2) - Q\D, \T = Q$. In this case, $\t_i$ is determined by
\be\label{xxy}
\t_i = \t - \s_{\t}\sqrt{Q} \int^{r}_{r_i} \f{dr}{\sqrt{ R }}\, ,
\ee
where $\s_{\t}$ dose not change its sign, as $\T$ is a constant. The expression of $\t_i$ indicates that the Carter constant $Q$ governs the bending of geodesics, which is expected to be small when describing the accretion flows around black holes. For small $Q$, we have $R \approx 2r(r^2+a^2) $, and Eq.~\eqref{xxy} can be expressed in terms of hypergeometric functions,
\bea\label{ti}
\t_i = \t - \s_{\t}\sqrt{\f{2Q}{a}} \bigg[ \sqrt{\f{r}{a}}\, \,_2F_1\bigg(\f{1}{4}, \f{1}{2}, \f{5}{4}, -\f{r^2}{a^2}\bigg) - \sqrt{\f{r_i}{a}}\, \,_2F_1\bigg(\f{1}{4}, \f{1}{2}, \f{5}{4}, -\f{r_i^2}{a^2}\bigg)   \bigg] \, .
\eea

\section{Magnetic field components in different frames}\label{AppC}

In the cases that both the spacetime and electromagnetic field are stationary and axisymmetric, and the fluid fulfills the ideal MHD condition of $F_{\mu\nu}u^\nu = 0$, the magnetic field 4-vector takes 
\bea\label{genB}
B^\m = \f{1}{\sqrt{-g}} \f{\pl_{\t}A_{\p}}{u^r} \big( u^\m u_t +\delta^\m_t \big) \, , 
\quad B^2 = \f{1}{-g}\bigg( \f{\pl_{\t}A_{\p}}{u^r}  \bigg)^2 \big( g_{tt} + u^2_t \big) \, .
\eea
Then, it is convenient to introduce the frames of ZAMOs as
\bea
\hat{e}^{\, \m}_{(t)} = \f{1}{\a} (\pl_t^{\, \m} + \o^{\p} \pl_{\p}^{\, \m}) \, ,\quad  \hat{e}^{\, \m}_{(i)} = \f{1}{\sqrt{g_{ii}}} \pl_i^{\, \m} \,, \quad  i = r,\t,\p \, 
\eea
with $\a $ being the lapse function, $\a  = \sqrt{-g_{tt} + \f{g^2_{t\p}}{g_{\p\p}}}$. The angular velocity of frame dragging is denoted as $\o^\p = -g_{t\p}/g_{\p\p}$.  The magnetic field component expressed in the tetrad of ZAMOs is given by $\hat{B}^{(a)} = \hat{e}^{(a)}_\m B^\m $, which leads to
\bea\label{BZAMO}
&&\hat{B}^{(t)} = \f{ \pl_\t A_\p}{\sqrt{-g}u^r} \bigg[ \a - \f{u_t}{\a}(u_t+\o^\p u_\p) \bigg]\, , \nn \\
&&\hat{B}^{(i)} =  \f{ \pl_\t A_\p}{\sqrt{-g}u^r} \sqrt{g_{ii}} \,  u^i u_t\, , \quad  i = r,\t, \, \nn \\
&&\hat{B}^{(\p)} =\f{ \pl_\t A_\p}{\sqrt{-g}u^r} \f{u_t u_\p + g_{t\p}}{\sqrt{g_{\p\p}}} \, .
\eea

Next, we explore the magnetic field components measured in the comoving frame of the fluid in the conical solution. Since $u^\t = 0$, the tetrad of the rest frame of the fluid can be chosen as
\bea\label{FRF}
&&s^\m_{(0)} =u^\m= \hat{u}^{(t)} \hat{e}^\m_{(t)}+\hat{u}^{(r)} \hat{e}^\m_{(r)}+\hat{u}^{(\p)} \hat{e}^\m_{(\p)} \, , \nn \\
&&s^\m_{(1)} =\f{ \hat{u}^{(\p)}}{\hat{u}}\hat{e}^\m_{(r)}- \f{\hat{u}^{(r)} }{\hat{u}}\hat{e}^\m_{(\p)}  \, , \nn \\
&&s^\m_{(2)} = \hat{e}^\m_{(\t)}  \, , \nn \\
&&s^\m_{(3)} = \hat{u} \, \hat{e}^\m_{(t)}+\f{\hat{u}^{(t)}\hat{u}^{(r)}}{\hat{u}} \hat{e}^\m_{(r)}+\f{\hat{u}^{(t)}\hat{u}^{(\p)}}{\hat{u}} \hat{e}^\m_{(\p)} \, ,
\eea
where $\hat{u}^{(a)} = \hat{e}^{(a)}_\m u^\m $ represents the 4-velocity measured by ZAMOs, and $\hat{u} = \sqrt{(u^{(r)})^2+(u^{(\p)})^2} = \sqrt{-1 + (\a u^t)^2}$ is the magnitude of the 4-velocity. Clearly, in the perspective of ZAMOs, $s^\m_{(1)}, s^\m_{(3)}$ is orthogonal and parallel to the fluid velocity, respectively. Then, the magnetic field components in the comoving frame are obtained through $B_{(a)} = s^\m_{(a)} B_\m$. Our calculations reveal that $B_{(a)} $ has the following simple form
\bea\label{Bcon}
B^{(0)} = B^{(2)} = 0 \, , \quad B^{(1)} = \f{ \pl_\t A_\p}{\hat{u}\sqrt{-g}}  \sqrt{g_{rr}g_{\p\p}}\, \o^\p \, , \quad  B^{(3)}  =  -\a  \f{ \pl_\t A_\p}{u^r\hat{u}\sqrt{-g}}  (u^tu_t +1) \,.
\eea 

In a Kerr spacetime, the magnetic field component Eq.~\eqref{genB} in the Boyer-Lindquist coordinates diverges as $r \rightarrow r_h$, while $B^2$ remains finite. In the comoving frame, it can be checked from Eqs. (\ref{u}), (\ref{Bcon}) that $B^{(a)}$ is finite outside the horizon.

\bibliographystyle{utphys}
\bibliography{note}
		
\end{document}